\documentclass[twocolumn,showpacs,preprintnumbers,amsmath,amssymb]{revtex4-1}
\usepackage{amssymb}
\usepackage{amsmath}
\usepackage{graphicx}
\usepackage{color}
\usepackage{hyperref}

\newcommand{\be}{\begin{equation}}
\newcommand{\ee}{\end{equation}}
\newcommand{\bea}{\begin{eqnarray}}
\newcommand{\eea}{\end{eqnarray}}
\newcommand{\beal}{\begin{aligned}}
\newcommand{\eeal}{\end{aligned}}

\usepackage{color}

\begin{document}

% \preprint{DCPT-16/15}


\title{On the thermodynamic phase structure of conformal gravity}


%\thanks{}
\author{Hao Xu}
\email{xuh5@sustc.edu.cn}

\affiliation{Department of Physics, Southern University of Science and Technology, Shenzhen 518055, China}

\author{Man-Hong Yung}
\affiliation{Institute for Quantum Science and Engineering and Department of Physics, Southern University of Science and Technology, Shenzhen 518055, China}
\affiliation{Shenzhen Key Laboratory of Quantum Science and Engineering, Shenzhen, 518055, China}


\date{\today}

\begin{abstract}
The solutions of the Einstein equation are a subset of the solutions of conformal (Weyl) gravity, but the difference from the action means that the black hole thermodynamics of the two gravity theories would be different. In this paper we explore the thermodynamic phase structure for the conformal gravity in the four-dimensional AdS space-time. Special emphasis is put on the dependence on the parameter $c_1$ in the linear-$r$ term in the metric. The thermodynamic phase structure of the conformal gravity is very rich, including two branches of equations of states, negative thermodynamic volume, zeroth-order phase transition, and Hawking-Page like phase transition.
\end{abstract}

% \pacs{04.70.-s, 04.70.Dy}

\maketitle

Gravity theories containing higher-order derivative terms in the action are of fundamental interest for various reasons, in particular, generalizing Einstein gravity. One of the possible choices of such theory is the Lovelock gravity \cite{Lovelock:1971yv}, where the action contains a sum of dimensionally-extended Euler densities and coincides with Einstein gravity in three and four dimensions. The integral of the $k$-th order term in Lovelock gravity gives the Euler character in dimension $d=2k$, and the Einstein-Hilbert action is precisely the first-order term. The equations of motion, which depend only on the Riemann tensor and not on its derivatives, are also ``Einstein-like''. Furthermore, the higher-order derivative terms are also of interest in the context of the AdS/CFT correspondence \cite{Maldacena:1997re,Witten:1998qj}. They arise naturally in string theory \cite{Boulware:1985wk}.

Since the $k$-th order term in the Lovelock gravity only affects the local geometry of the manifold in $d\geq 2k+1$, all the higher-order terms($k\geq 2$) do not change the Einstein manifold in the lower dimensions. However, the higher-order  terms do play an important role in the renormalization theory \cite{Stelle:1976gc}. One of the gravity theories, known as the conformal (Weyl) gravity, which is described by a pure Weyl squared action, has been shown to be perturbatively renormalizable in four dimensions, although the massive modes are ghostlike. It is a local gauge theory of the conformal group, so the equation of motion determines the metric only up to an arbitrary conformal factor. The solutions of the Einstein equation are a subset of the solutions of conformal gravity; any space-time that is conformal to an Einstein space arises naturally as a solution of conformal gravity. In 2011, Maldacena demonstrated a remarkable result showing that conformal gravity with a Neumann boundary condition can select the Einstein solution out of conformal gravity \cite{Maldacena:2011mk,Anastasiou:2016jix}.

Conformal gravity is also of interest for cosmology. Although Einstein gravity can well describe the physics within the solar system, such as the gravitational bending of light and the precession of the perihelion of Mercury, there are still some puzzles left on scales far beyond the distances of the Solar system. For example, the galactic rotation curves are not consistent with the predictions of Einstein's gravity;  the unknown ``dark matter", which interacts with the baryonic matter only via gravity or via the weak force, is introduced to fix the problem. In addition, the concept of  dark energy is also introduced, which is much larger than the zero-point energy of the matter fields, in order to provide the energy source to explain the observation of our accelerating universe.

Alternatively, one may also question whether it is possible to modify the gravity theory in order to explain the physics at a large scale, while maintaining the behavior at the scale of the Solar system. Since conformal gravity possesses more solutions than Einstein gravity, the dark matter, dark energy problems can be fixed within the gravity theory \cite{Mannheim:1988dj,Mannheim:2005bfa,Mannheim:2010ti,Mannheim:2011ds}, which motivates us to  investigate conformal gravity in more detail. In particular, although conformal gravity and Einstein gravity may share the same space-time solutions, the difference from the action means that the black hole thermodynamics of the two gravity theories would be different, which points to an area where extensive research should be carried out.

Here, we focus on the thermodynamic phase structure of the conformal gravity in the four-dimensional AdS space-time. Moreover, we do not assume the  ``physical constants", such as Yukawa coupling constants, cosmological constant, and Lovelock coefficients, to be fixed values; they may be dynamical variables resulting form the vacuum expectation values; it is therefore reasonable to include them into the thermodynamic laws \cite{Gibbons:1996af,Kastor:2010gq,Xu:2013zea,Anabalon:2015xvl,Astefanesei:2018vga}. Recently, the idea of including the cosmological constant in the first law of black hole thermodynamics becomes popular. See, e.g., \cite{Kastor:2009wy,Kubiznak:2012wp,Mann:2015luq,Gregory:2017sor,Gunasekaran:2012dq,Cai:2013qga,
Altamirano:2013uqa,Zou:2013owa,Altamirano:2014tva,Wei:2014hba,Johnson:2014yja,Xu:2014tja,Frassino:2014pha,
Dolan:2014vba,Xu:2015rfa,Majhi:2016txt,Bhattacharya:2017hfj,Bhattacharya:2017nru,Xu:2018fag} for references and reviews. In particular, in AdS space-time the negative cosmological constant behaves like the pressure, while its conjugate variable can be considered as a thermodynamic volume. However, in Einstein gravity, the cosmological constant $\Lambda$ comes from the action, thus varying $\Lambda$ requires changing the system. Remarkably, one do not have to worry about this problem in conformal gravity, as $\Lambda$ arises as the integral constant of the solution \cite{Lu:2012xu,Xu:2014kwa,Xu:2017ahm} instead of the action. This makes the analysis for the phase structure self-contained.

We start by giving a brief review of the black hole thermodynamics of conformal gravity \cite{Lu:2012xu}. First, the action is given by a square of the Weyl tensor,
\begin{align}
S=\alpha\int \mathrm{d}^4x \sqrt{-g}C^{\mu\nu\rho\sigma}
C_{\mu\nu\rho\sigma} \ .
\label{action}
\end{align}
The coupling constant $\alpha$ plays an important role in critical gravity \cite{Bergshoeff:2011ri,Alishahiha:2011yb,Gullu:2011sj}. However, here it is independent of the equation of motion and do not change any qualitative feature of the thermodynamic quantities. Without loss of generality, we set $\alpha=1$ in the following. Moreover, the equation of motion is fourth order,
\begin{align}
(2\nabla^\rho \nabla^\sigma + R^{\rho\sigma})
C_{\mu\rho\sigma\nu}=0 \,.
\end{align}
The most general spherical black hole solution for conformal gravity takes the following form \cite{Riegert:1984zz,Klemm:1998kf}
\begin{align}
  &\mathrm{d} s^2=-f(r)\mathrm{d}t^2+\frac{\mathrm{d}r^2}{f(r)}+r^2\mathrm{d}
  \Omega_{S^2}^2, \label{metric}
\end{align}
where $\mathrm{d}\Omega_{S^2}^2$ is the line element of a 2-dimensional sphere and
\begin{align}
  f(r)=c_0+c_1 r+\frac{d}{r}-\frac{1}{3}\Lambda r^2.
\end{align}
Due to the conformal symmetry of the action, the Weyl rescaling of the above metric remains a static spherically symmetric solution, so the Birkhoff's theorem only restricts the static spherically symmetric solutions to a conformal class. There are four different integral constants, $c_0, c_1, d, \Lambda$ in the metric. Three of them must obey a constraint,
\begin{align}
  c_0^2=3c_1d+1.
\label{relation}
\end{align}
When $c_1=0$, this solution reduces to the well-known Schwarzschild (A)dS space-time. Notice that there exists a discrete freedom in choosing the constant $c_0$,
\begin{align}
c_0=\pm\sqrt{3c_1d+1}.
\label{c0}
\end{align}
Mathematically, we can also solve $c_1$ in terms of $c_0$ and $d$. However, from the thermodynamical perspective, it is preferable to take $c_1$ as independent thermodynamical parameter \cite{Lu:2012xu}.

In our case, $\Lambda$ plays the role of the cosmological constant. It is an integral constant but not from the action. The energy, which is defined by the conserved charge of the timelike killing vector, should be identified as enthalpy $H$ of the system \cite{Lu:2012xu}
\begin{align}
  H=\,{\frac {\left( c_{{1}}c_{{0}}-c_{{1}}-16\pi P\,d
 \right) }{12\pi }}.
\end{align}
The other thermodynamic quantities can also be obtained. The temperature is proportional to the surface gravity at the horizon with radius $r_0$
\begin{align}
  T=\,{\frac {8\pi P\,r_0^{3}-3\,c_{{0}}r_{{0}}-6\,d}{12\pi \,{r
_{{0}}}^{2}}},
\label{temperature}
\end{align}
where $r_0$ is the largest root of $f(r_0)=0$ in AdS space-time, and its conjugate, i.e. the entropy is \cite{Lu:2012xu}
\begin{align}
  S=\,{\frac {\,\left( r_0-c_{{0}}r_{{0}}-\,  3\,
d \right) }{3r_{{0}}}},
\label{S}
\end{align}
which is a function of $c_0$, $r_0$ and $d$ rather than being proportional to the area of the horizon. If we also treat $c_1$ as a variable, its conjugate quantity is
\begin{align}
  \Psi=\,{\frac {\, \left( c_{{0}}-1 \right) }{12\pi }}.
\end{align}
We take the cosmological constant $\Lambda$ as the pressure,
\begin{align}
  P=-\frac{\Lambda}{8\pi},
\end{align}
then the thermodynamic volume is
\begin{align}
  V=\left(\frac{\partial H}{\partial P}\right)_{S,c_1}
  =-\,{\frac {2d}{3 }}.
  \label{V}
\end{align}

So we can get the first law of black hole thermodynamics in conformal gravity
\begin{align}
  {\it \mathrm{d}H}=T{\it \mathrm{d}S}+ \Psi\,
  \mathrm{d}c_1 + V\,\mathrm{d}P
\end{align}
and the Smarr relation \cite{Lu:2012xu}
\begin{align}
  H=2PV+\Psi\,c_1 \label{Smarr}.
\end{align}
The Gibbs free energy can be obtained by the relation (or by using the Euclidean action)
\begin{align}
G=H-TS=\frac{2(c_0-1)r_0+(3+8\pi Pr_0^2)d}{12\pi r_0^2}.
\end{align}

Notice that here we have fixed the parameter $\alpha$, which has dimensions of [length]$^2$. If we consider the contributions of the $\alpha$, we will have $H\rightarrow\alpha H$, $S\rightarrow\alpha S$, $V\rightarrow\alpha V$, $\Psi\rightarrow\alpha \Psi$, $G\rightarrow\alpha G$ and obtain the ordinary scaling dimensions of these thermodynamic quantities \cite{Lu:2012xu}. The $\alpha$ does not change any qualitative feature of the thermodynamic quantities or the Smarr relation.

To study the phase structure of conformal gravity, we should begin with the equations of state in $P-V$ plane. By using $f(r_0)=0$, \eqref{relation}, and \eqref{temperature} to eliminate other unnecessary coefficients, we have two solutions of $P(T,r_0)$ and $V(T,r_0)$. The first one is
\begin{align}
\begin{split}
  &P_1=\,{\frac {T}{2r_{{0}}}}-\,{\frac {c_1r_{{0}}- \,\sqrt
{1-4\,\pi \,T c_1 r_0^{2}}}{8{\pi }
{r_{{0}}}^{2}}},\\
  &V_1=\frac{2r_0}{9}\bigg(4\pi Tr_0-c_1 r_0-2\sqrt{1-4\pi T c_1 r_0^2}\bigg),
\label{solution1}
\end{split}
\end{align}
while the other is
\begin{align}
\begin{split}
  &P_2=\,{\frac {T}{2r_{{0}}}}-\,{\frac {c_1r_{{0}}+ \,\sqrt
{1-4\,\pi T c_1 r_0^{2}}}{8{\pi }
{r_{{0}}}^{2}}},\\
  &V_2=\frac{2r_0}{9}\bigg(4\pi Tr_0-c_1 r_0+2\sqrt{1-4\pi  T c_1 r_0^2}\bigg).
\label{solution2}
\end{split}
\end{align}
Here we treat $r_0$ as an intermediate parameter to avoid the overwhelming complexity of formula $P(T,V)$. The possible critical behavior of the black hole can be found by using following two equations,
\begin{align}
\frac{\partial P}{\partial V}=0,\quad \frac{\partial^2 P}{\partial V^2}=0.
\end{align}
The only physical solution, which satisfies $P>0$, $r_0>0$ and $T>0$, comes from the second equations of state $(P_2,V_2)$, if and only if $c_1>0$. The critical point reads
\begin{align}
\begin{split}
&P_c=\frac{(2-3X^2-\sqrt{3}X)c_1^2}{24\pi X^3},\quad T_c=\frac{c_1}{6\pi X^2}\\
&V_c=\frac{4-6X^2+4\sqrt{3}X}{27c_1},
\end{split}
\end{align}
where $X=\sqrt{2}-2\sqrt{3}/{3}$. Since the qualitative features of the phase structure depend on the sign of $c_1$, we will investigate them in three different situations, $c_1=0$, $c_1>0$ and $c_1<0$.
\\

1. $c_1=0$. Firstly we consider the case of $c_1=0$. In this case, the metric reduces to the well-known Schwarzschild AdS black hole. However, due to the difference from the action, the phase structure of the conformal gravity can be different from the Einstein gravity. Now the equations of states for the two
solutions become
\begin{align}
P_1=\frac{T}{2r_0}+\frac{1}{8\pi r_0^2},\quad V_1=\frac{8\pi T}{9}r_0^2-\frac{4r_0}{9},
\end{align}
and
\begin{align}
P_2=\frac{T}{2r_0}-\frac{1}{8\pi r_0^2},\quad V_2=\frac{8\pi T}{9}r_0^2+\frac{4r_0}{9},
\end{align}
which corresponds to $c_0=-1$ and $c_0=1$ respectively.

\begin{figure}
\begin{center}
\includegraphics[width=0.3\textwidth]{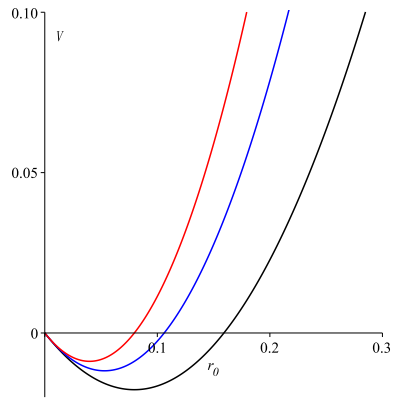}
\caption{The isothermal curves in $V-r_0$ plane for $c_1=0$ and $c_0=-1$. The temperature of the isotherms decreases from left to right.}
\label{fig1}
\end{center}
\end{figure}

Before we investigate the phase structure of the black hole in $(P, V) $ plane, we give a brief discussion of the thermodynamic volume $V_1$. From the above formula of $V_1$, we can find that $V_1\rightarrow 0$ as $r_0\rightarrow 0$, and $V_1\rightarrow +\infty$ as $r_0\rightarrow +\infty$. This is normal. However, $V_1$ also admits a minimum negative value. From the black hole thermodynamics and general relativity point of view, this negative volume does have physics meaning. In Einstein gravity, the thermodynamic volume of the black hole is the flat space volume of a ball whose radius is the horizon. Creating the black hole by adding energy to the pure AdS space-time involves cutting out the volume $V$. On the other hand, the negative volume was firstly noted in Taub-NUT case \cite{Johnson:2014xza}. Instead of cutting out volume, environment has to do work on the system to create the solution while the universe has to increase its volume. From Fig.\ref{fig1} we can see there is a region of $r_0$ where the thermodynamic volume takes the negative value $V_1$. This also means the reverse isoperimetric inequality \cite{CveticEtal:2011} is violated and the black hole is super-entropic \cite{Hennigar:2014cfa}.

\begin{figure}
\begin{center}
\includegraphics[width=0.3\textwidth]{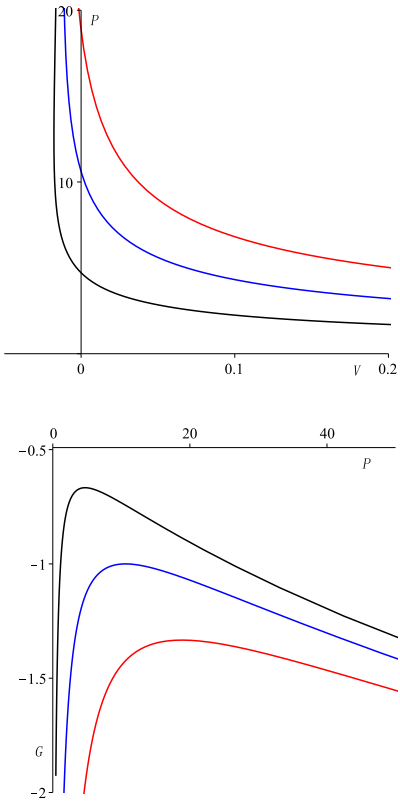}
\includegraphics[width=0.3\textwidth]{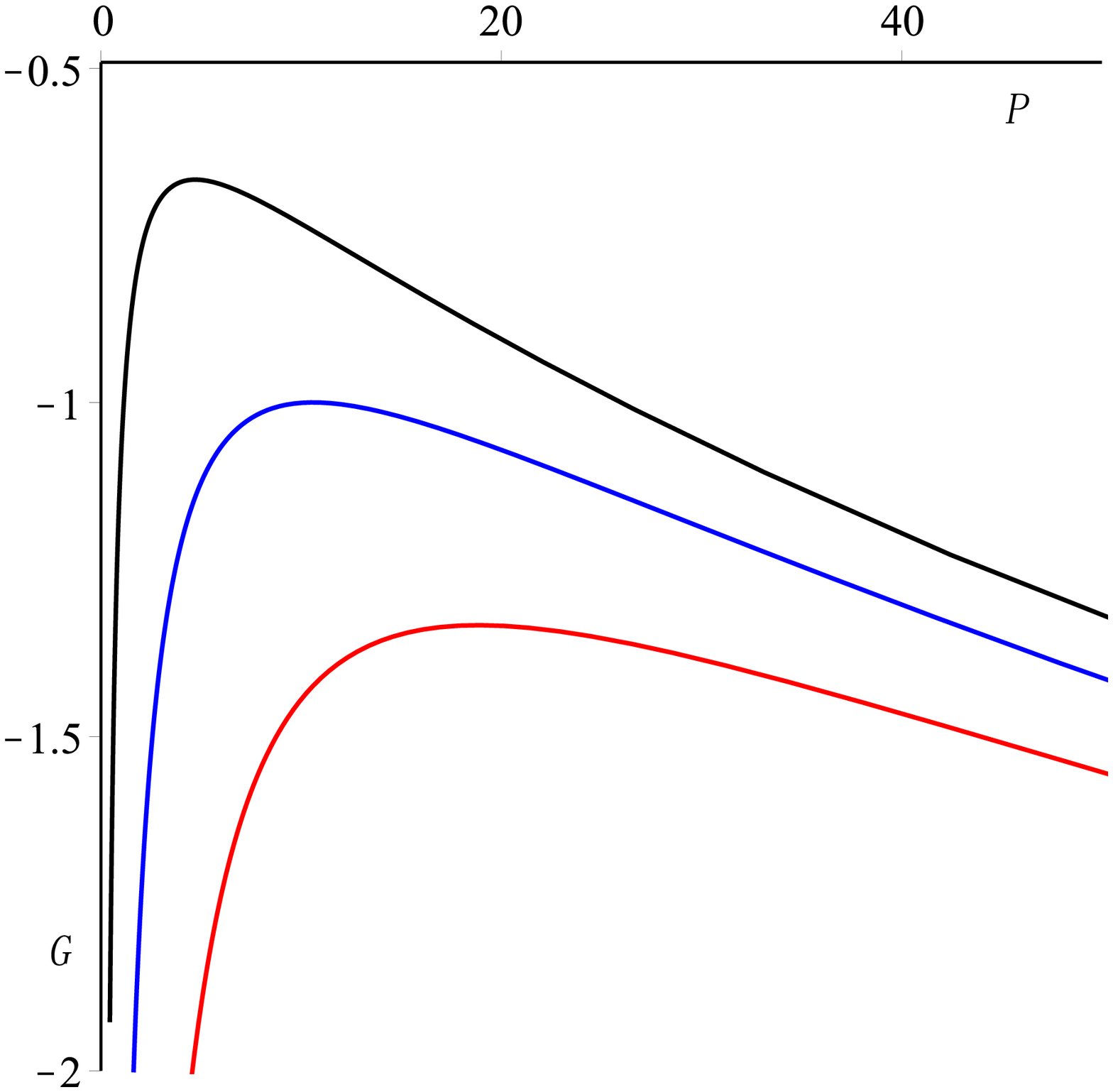}
\caption{The isothermal curves in $P-V$ plane and $G-P$ plane for $c_1=0$ and $c_0=-1$. The temperature decreases/increases from top to bottom in the $P-V$/$G-P$ plane.}
\label{fig2}
\end{center}
\end{figure}

We investigate the phase structure of $(P_1,V_1)$ in Fig.\ref{fig2}. There is always a region of isotherms with the negative volume in the $P-V$ plane. The isotherms in $G-P$ plane are also non-monotonic, predicting the reverse of the sign of $V$.

\begin{figure}
\begin{center}
\includegraphics[width=0.3\textwidth]{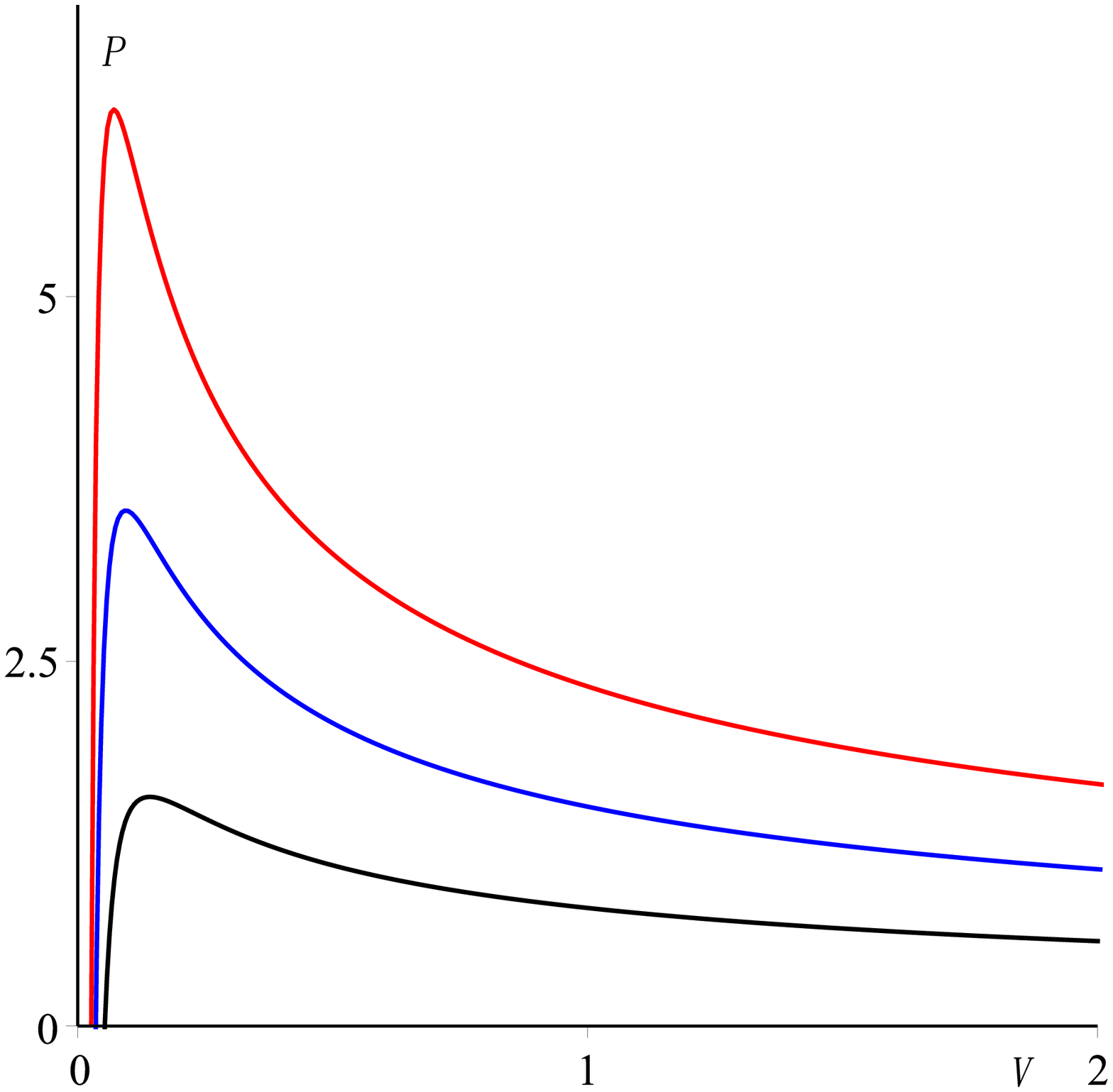}
\includegraphics[width=0.3\textwidth]{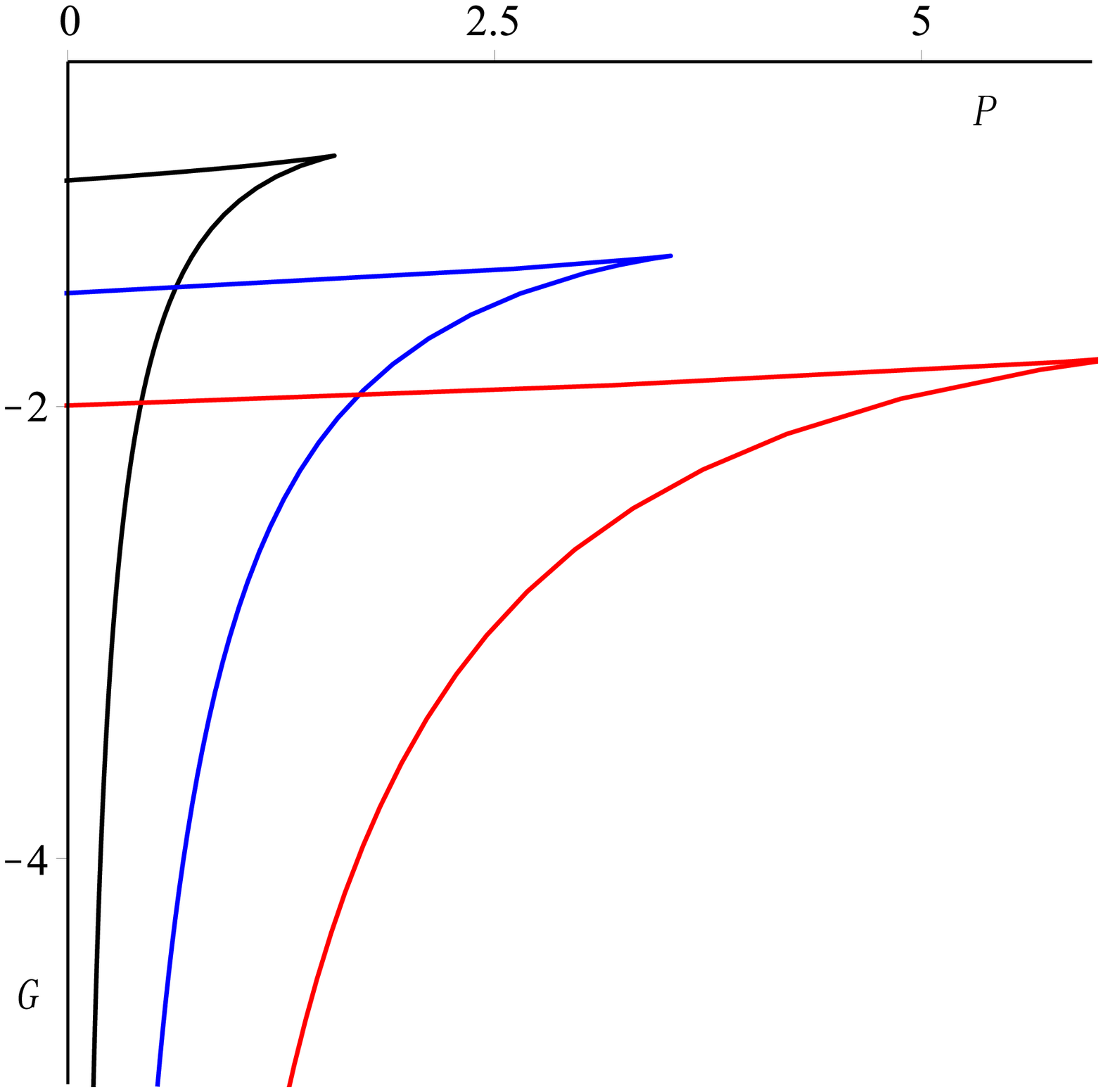}
\caption{The isothermal curves in $P-V$ plane and $G-P$ plane for $c_1=0$ and $c_0=1$. The temperature decreases/increases from top to bottom in the $P-V$/$G-P$ plane.}
\label{figadd}
\end{center}
\end{figure}

Then we consider the case of $(P_2,V_2)$. The thermodynamic volume $V_2$ are always positive in $r_0>0$. In Fig.\ref{figadd} we present the isotherms in $P-V$ and $G-P$ planes of $(P_2,V_2)$. The black hole has an analogy with the Hawking-Page phase transition in Schwarzschild AdS black hole from Einstein gravity \cite{Hawking:1982dh}. From the $P-V$ plane we can find for each isotherm there is a maximum, which precisely corresponds to the minimal black hole radius for a given temperature. The curve with positive slope to the left of the maximum corresponds to small black holes that are thermodynamically unstable and can not be in thermal equilibrium with a thermal bath of radiation, whereas the curve with negative slope corresponds to large black holes that are locally thermodynamically stable. In the $G-P$ plane the lower thermodynamically preferred branch corresponds to the large stable black holes.

2. $c_1>0$. There are obvious differences in the phase structure between $c_1>0$ and $c_1<0$, but the exact value of $c_1$ does not change the qualitative features. For simplicity we will set $c=1$ for the case $c_1>0$ and $c=-1$ for $c_1<0$. In $c_1=1$ the equations of states for the two solutions become
\begin{align}
\begin{split}
  &P_1=\,{\frac {T}{2r_{{0}}}}-\,{\frac {r_{{0}}- \,\sqrt
{1-4\,\pi \,T r_0^{2}}}{8{\pi }
{r_{{0}}}^{2}}},\\
  &V_1=\frac{2r_0}{9}\bigg(4\pi Tr_0-r_0-2\sqrt{1-4\pi Tr_0^2}\bigg),
\end{split}
\end{align}
and
\begin{align}
\begin{split}
  &P_2=\,{\frac {T}{2r_{{0}}}}-\,{\frac {r_{{0}}+ \,\sqrt
{1-4\,\pi T r_0^{2}}}{8{\pi }
{r_{{0}}}^{2}}},\\
  &V_2=\frac{2r_0}{9}\bigg(4\pi Tr_0-r_0+2\sqrt{1-4\pi  Tr_0^2}\bigg).
\end{split}
\end{align}
Due to the different sign of the $\sqrt{1-4\pi Tr_0^2}$ term, it is normal that we treat $(P_1,V_1)$ and $(P_2,V_2)$ as different branches of the solution. However, the different sign appears because we use the $r_0$ as the intermediate parameter. The key to distinguish the different solutions is the sign of $c_0$. Since $c_0$ serves to be of topological nature, which can only take $0,\pm 1$ in Einstein gravity, its sign cannot be changed by varying the dependent parameters $c_1$ and $d$. Inserting the formulas of $P_1$ and $V_1$ into $f(r_0)=0$, we can obtain
\begin{align}
c_0=-r_0-\sqrt{1-4\pi Tr_0^2},
\end{align}
which is always negative. However, if we insert $P_2$ and $V_2$ into $f(r_0)=0$,
\begin{align}
c_0=-r_0+\sqrt{1-4\pi Tr_0^2}.
\end{align}
If $1/\sqrt{16\pi^2 T^2-4\pi T+1}<r_0<1/\sqrt{1+4\pi T}$, $c_0>0$. The lower bound of $r_0$ is to secure the AdS background. However, if $1/\sqrt{1+4\pi T}<r_0<1/\sqrt{4\pi T}$, $c_0$ is negative, meaning $(P_2,V_2)$ belongs to the branch of $c_0<0$.

\begin{figure}
\begin{center}
\includegraphics[width=0.3\textwidth]{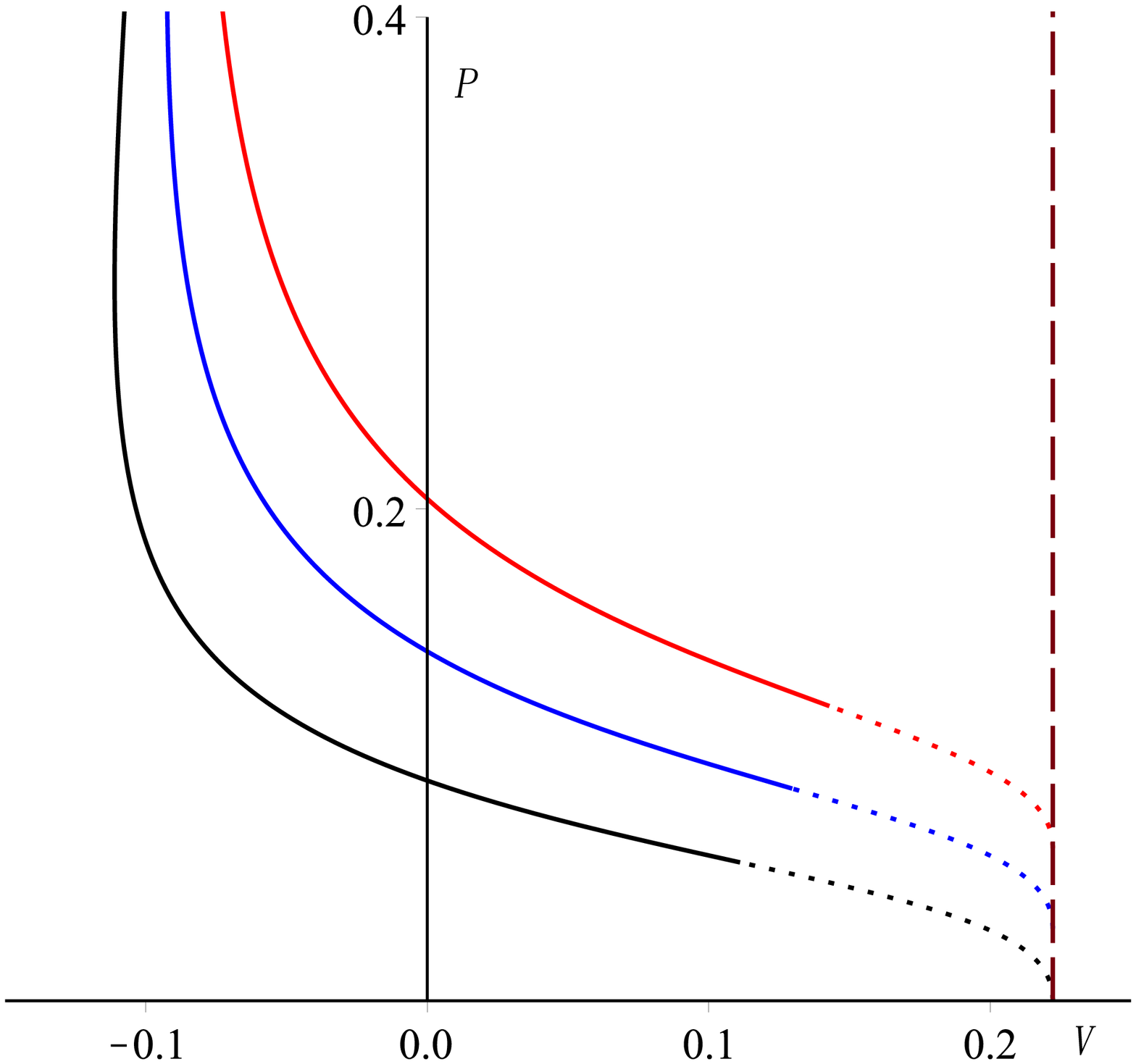}
\includegraphics[width=0.3\textwidth]{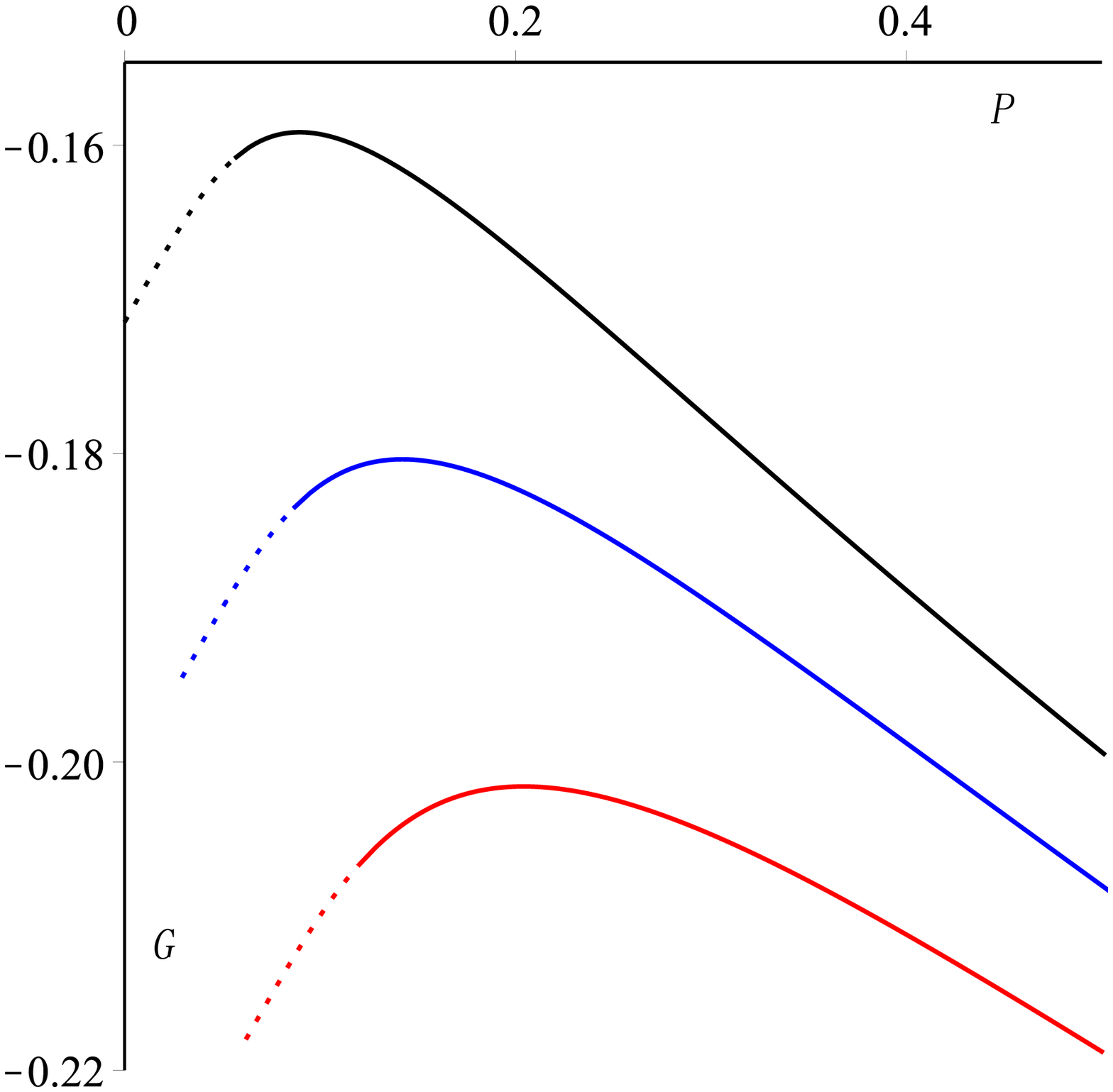}
\caption{The isothermal curves in $P-V$ plane and $G-P$ plane for $c_0<0$ and $c_1>0$. The solid and dotted lines correspond to the formulas of $(P_1,V_1)$ and $(P_2,V_2)$ respectively. The temperature decreases/increases from top to bottom in the $P-V$/$G-P$ plane.}
\label{fig3}
\end{center}
\end{figure}

In Fig.\ref{fig3} we present the isotherms in $P-V$ and $G-P$ planes as $c_0<0$. The solid and dotted lines correspond to the formulas of $(P_1,V_1)$ and $(P_2,V_2)$ respectively. There is also a region in the isotherms with the negative volume. From \eqref{relation}, \eqref{V} we know the volume has a upper bound, corresponding to $d=-\frac{1}{3c_1}$ and $c_0=0$. Each curve in the $G-P$ plane also admits a maximal value.

\begin{figure}
\begin{center}
\includegraphics[width=0.3\textwidth]{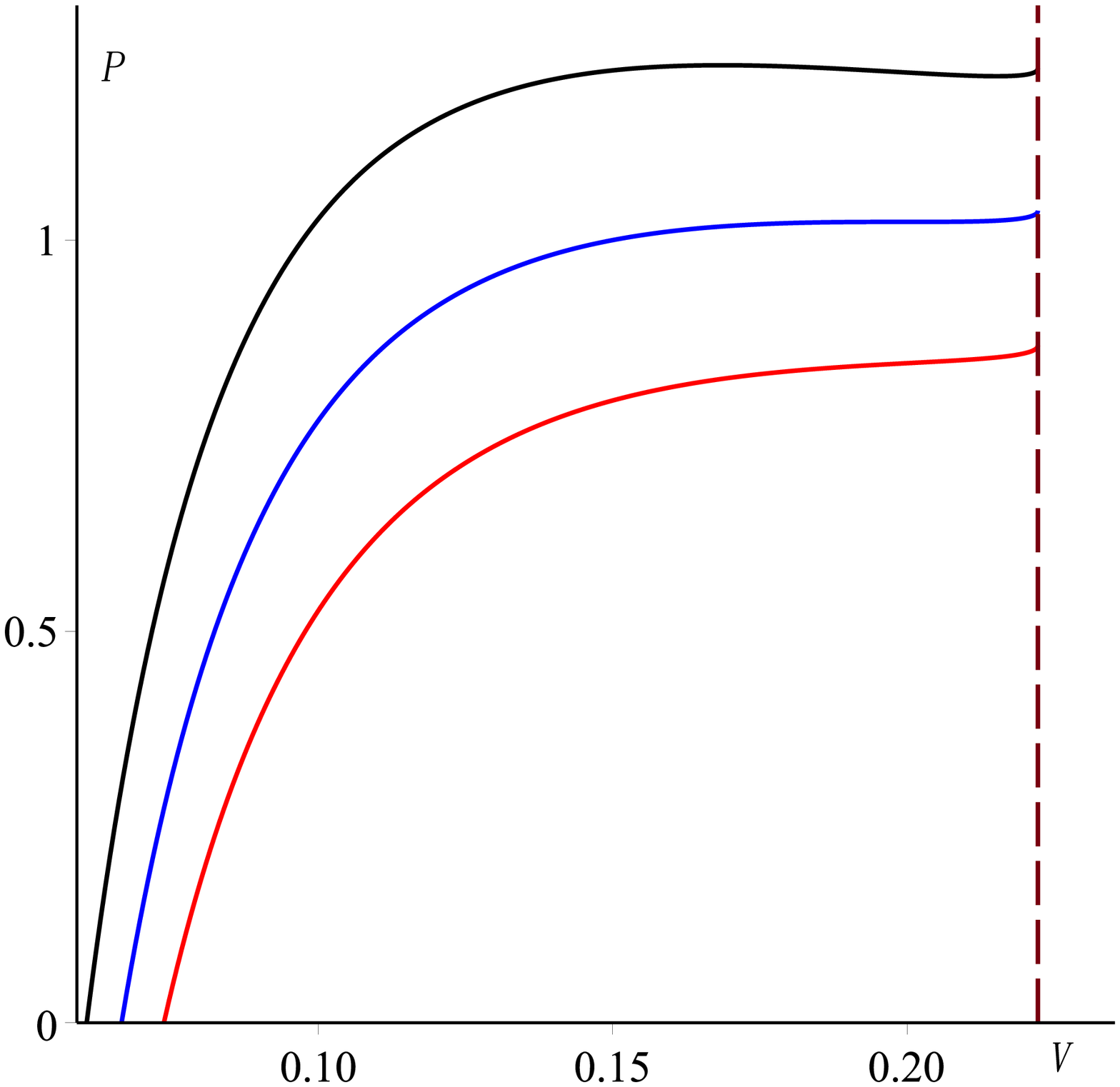}
\includegraphics[width=0.3\textwidth]{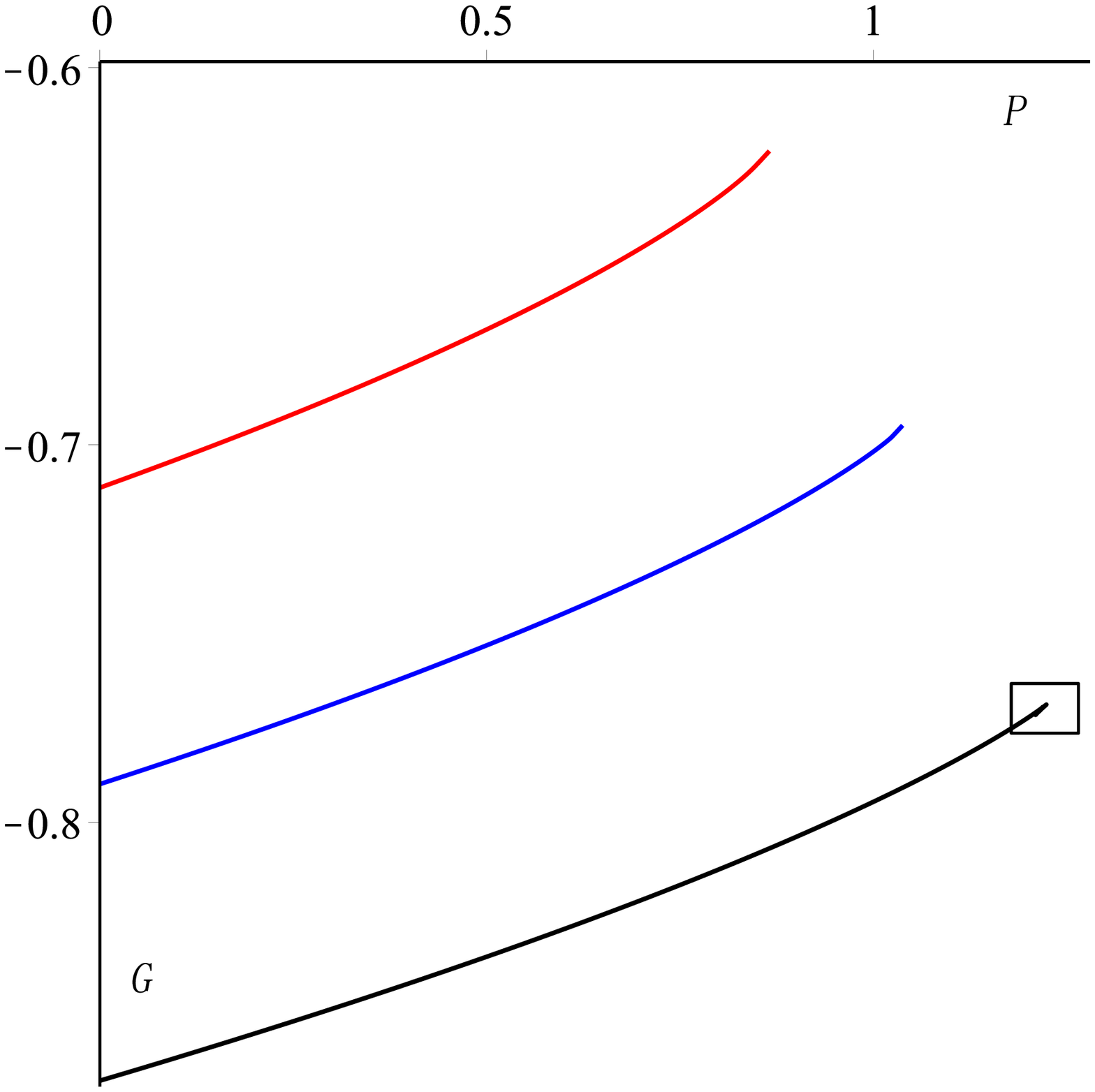}
\caption{The isothermal curves in $P-V$ plane and $G-P$ plane for $c_0>0$ and $c_1>0$. From top/bottom to bottom/top the temperature of the curves is $1.1T_c$, $T_c$, $0.9T_c$ in the $P-V$/$G-P$ plane respectively. There is a small unsmooth region at $T=1.1T_c$ in the $G-P$ plane.}
\label{fig4}
\end{center}
\end{figure}

\begin{figure}
\begin{center}
\includegraphics[width=0.3\textwidth]{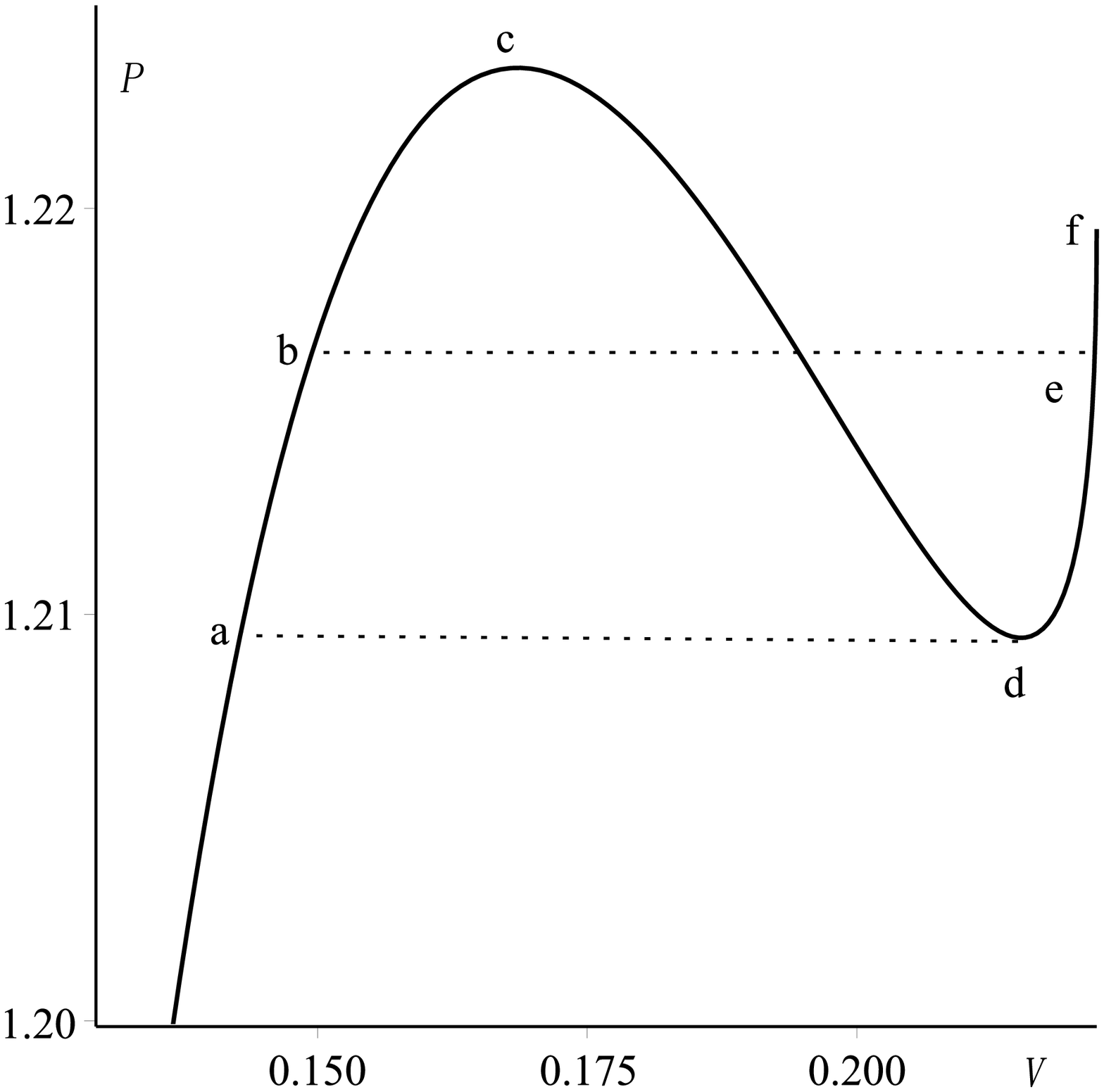}
\includegraphics[width=0.3\textwidth]{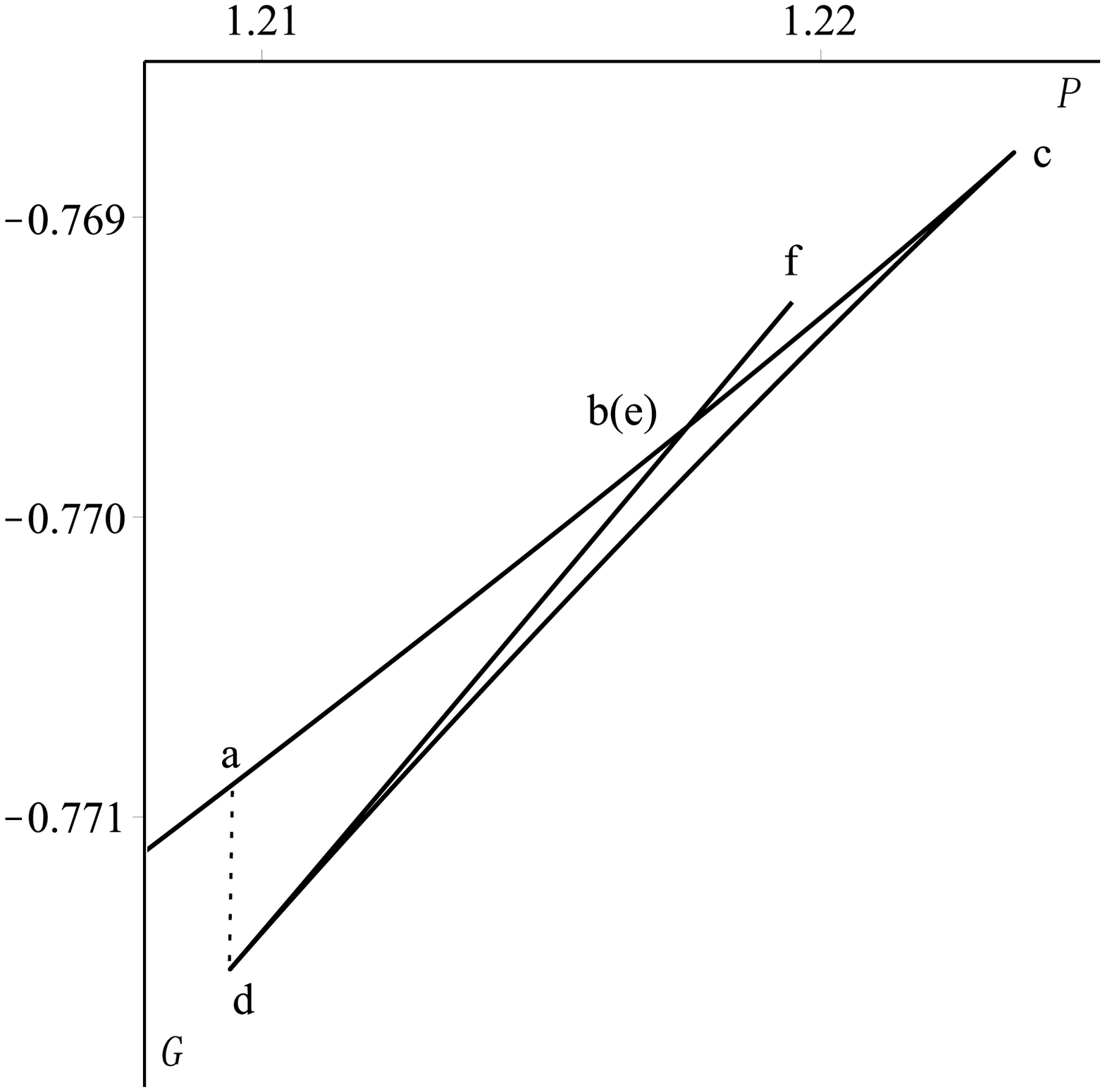}
\caption{Magnification of the unsmooth region in the $G-P$ plane and its corresponding part in the $P-V$ plane for $c_0>0$ and $c_1>0$.}
\label{fig5}
\end{center}
\end{figure}

Then we investigate the phase structure in $c_0>0$. Notice now the equations of states admit a critical point. In Fig.\ref{fig4} we present the isotherms in $P-V$ and $G-P$ planes. From top to bottom the temperature of the curves in the $P-V$ plane is $1.1T_c$, $T_c$, $0.9T_c$ respectively. In the $G-P$ plane we can observe there is a small unsmooth region in the $T=1.1T_c$ curve. We zoom in this area and the corresponding part of the $P-V$ plane in Fig.\ref{fig5}. The points marked on the two curves have one-to-one correspondence. In $P-V$ plane there is an oscillating part in the isotherm. Traditionally it should be replaced by an isobaric line determined by the Maxwell construction \cite{Xu:2015hba}, predicting a stable coexistence state of two different black holes and the first order phase transition. However, from the $G-P$ plane we can find this oscillating part corresponds to a \emph{reverse} swallow tail. The coexistence state b(e) is also unstable. There is one zeroth order phase transition, instead of the first order, at point a/d(there will be two if the pressure at f is larger than at c, depending on the chosen temperature). Unlike Einstein gravity, this phase transition can be found in $T>T_c$ rather than $T<T_c$.\\

3. $c_1<0$. Now we consider the case of $c_1<0$. As mentioned before, we set $c_1=-1$ for simplicity. The equations of states become
\begin{align}
\begin{split}
  &P_1=\,{\frac {T}{2r_{{0}}}}+\,{\frac {r_{{0}}+ \,\sqrt
{1+4\,\pi \,T r_0^{2}}}{8{\pi }
{r_{{0}}}^{2}}},\\
  &V_1=\frac{2r_0}{9}\bigg(4\pi Tr_0+ r_0-2\sqrt{1+4\pi T r_0^2}\bigg),
\end{split}
\end{align}
and
\begin{align}
\begin{split}
  &P_2=\,{\frac {T}{2r_{{0}}}}+\,{\frac {r_{{0}}- \,\sqrt
{1+4\,\pi T r_0^{2}}}{8{\pi }
{r_{{0}}}^{2}}},\\
  &V_2=\frac{2r_0}{9}\bigg(4\pi Tr_0+ r_0+2\sqrt{1+4\pi T r_0^2}\bigg).
\end{split}
\end{align}

Inserting the above formulas of $P_1$ and $V_1$ into $f(r_0)=0$, we have
\begin{align}
c_0=r_0-\sqrt{1+4\pi T r_0^2}.
\end{align}
If $T>1/4\pi$, $c_0$ is always negative. However, if $T<1/4\pi$, we can find $c_0<0$ as $0<r_0<1/\sqrt{1-4\pi T}$, and $c_0>0$ as $r_0>1/\sqrt{1-4\pi T}$.

Thus far we haven't discussed the black hole entropy, because it remains non-negative in all the choices of parameters as $c_1\geqq 0$. However, in $c_1=-1$, if we insert $P_1$ and $V_1$ into the formula of $S$, we have
\begin{align}
S=\frac{1}{3}+\frac{4\pi Tr_0}{3}-\frac{1}{3}\sqrt{1+4\pi T r_0^2}.
\end{align}
We can find $S$ is always positive as $T>1/4\pi$. If $T<1/4\pi$, $S$ is positive only in $0<r_0<2/(1-4\pi T)$.

Now let us make a short summary of our known results. If $T>1/4\pi$, $c_0<0$ and $S>0$ in any $r_0>0$. If $T<1/4\pi$, $c_0<0$, $S>0$ in $0<r_0<1/\sqrt{1-4\pi T}$, and $c_0>0$, $S>0$ in $1/\sqrt{1-4\pi T}<r_0<2/(1-4\pi T)$.

\begin{figure}
\begin{center}
\includegraphics[width=0.3\textwidth]{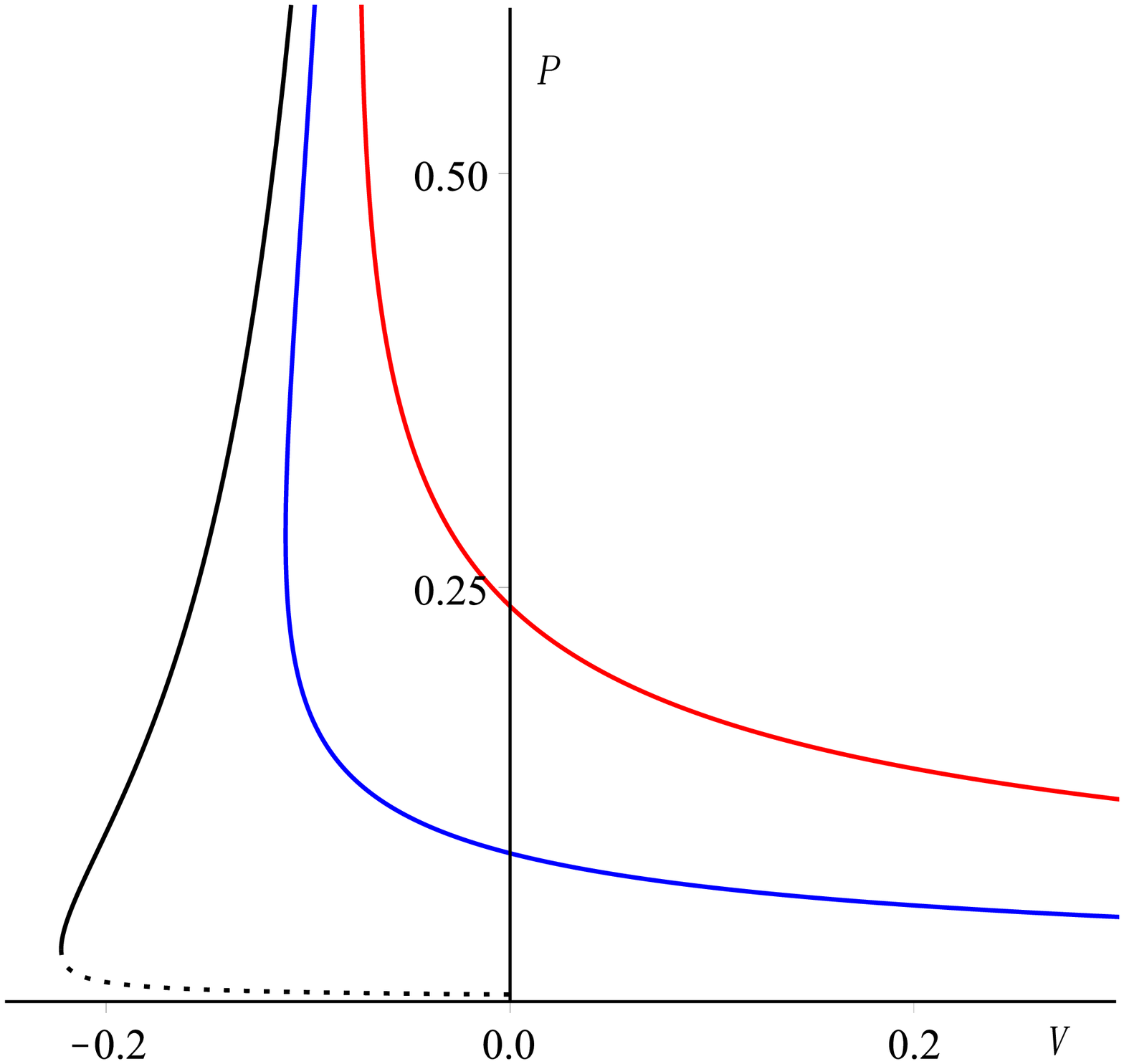}
\includegraphics[width=0.3\textwidth]{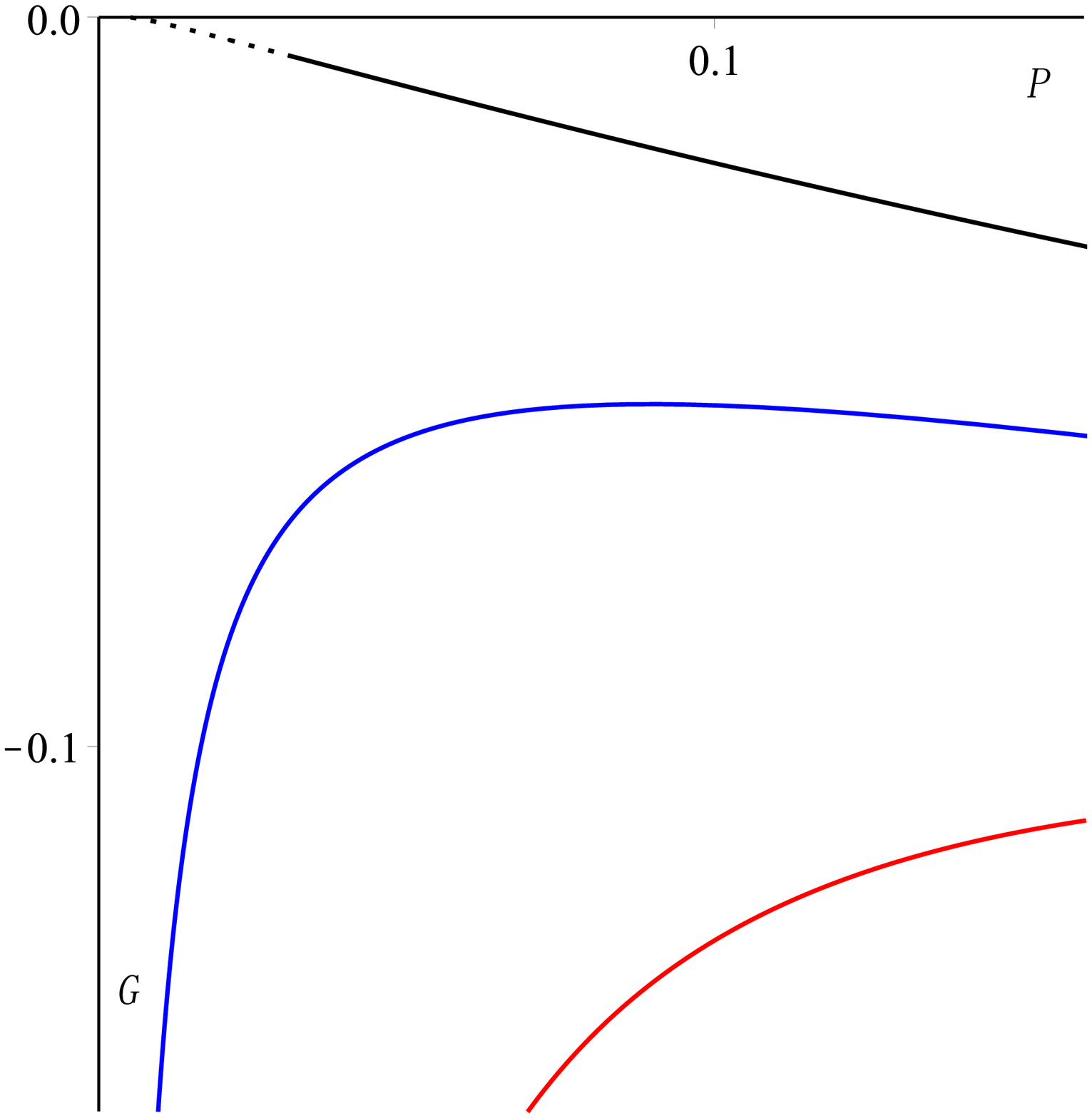}
\caption{The isothermal curves in $P-V$ plane and $G-P$ plane for $(P_1,V_1)$ in $c_1<0$. From top to bottom the temperature of the curves decreases/increases in the $P-V$/$G-P$ plane. The dotted part of the bottom/top line in the $P-V$/$G-P$ plane corresponds to the $c_0>0$ case.}
\label{fig6}
\end{center}
\end{figure}

In Fig.\ref{fig6} we present the isotherms in $P-V$ and $G-P$ planes of $(P_1,V_1)$. The phase structure is quite similar to the case of $c_1=0$. The only difference is at the bottom/top line($T<1/4\pi$) in the $P-V$/$G-P$ plane, where the dotted line corresponds to the $c_0>0$ case.

\begin{figure}
\begin{center}
\includegraphics[width=0.3\textwidth]{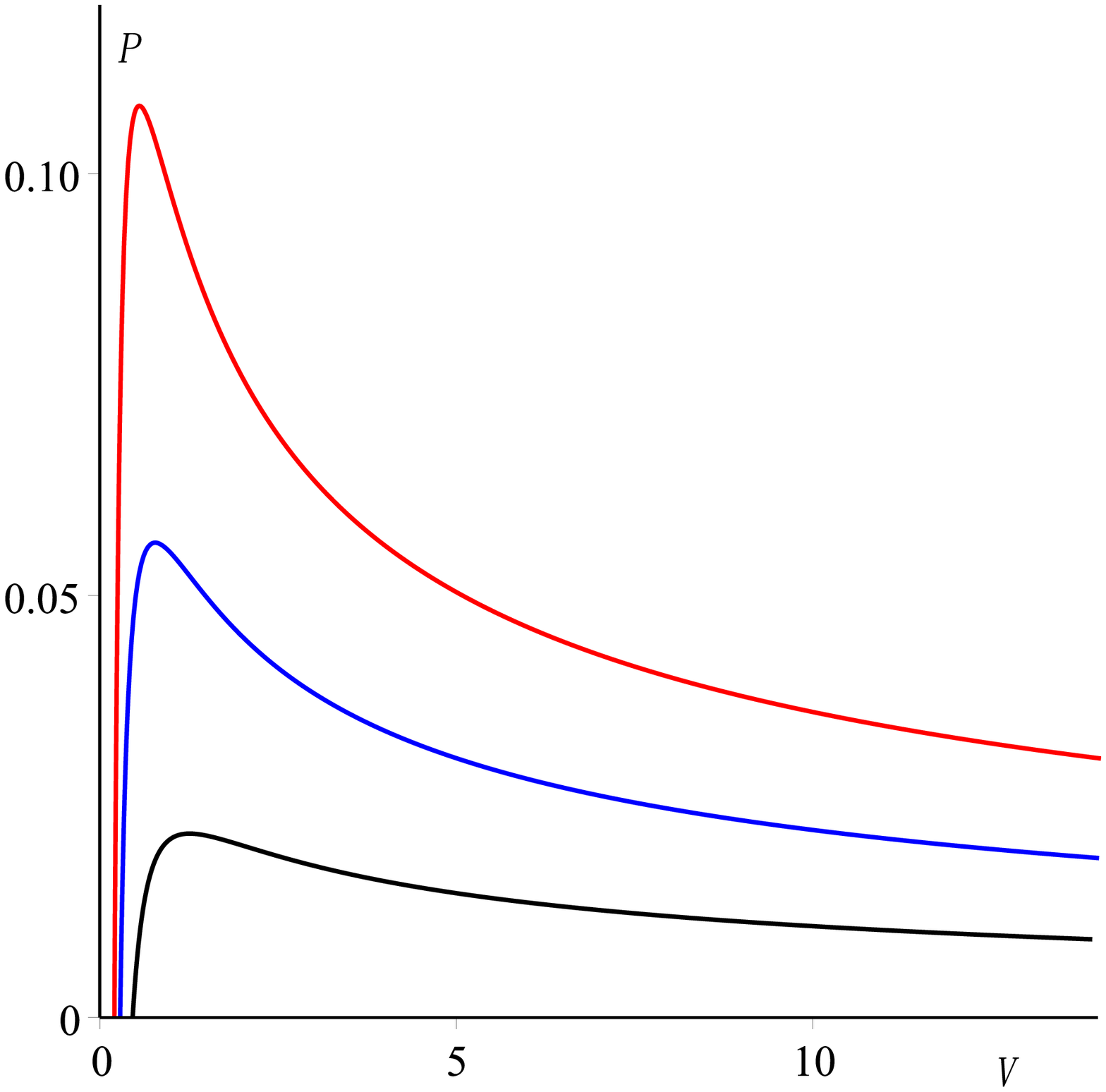}
\includegraphics[width=0.3\textwidth]{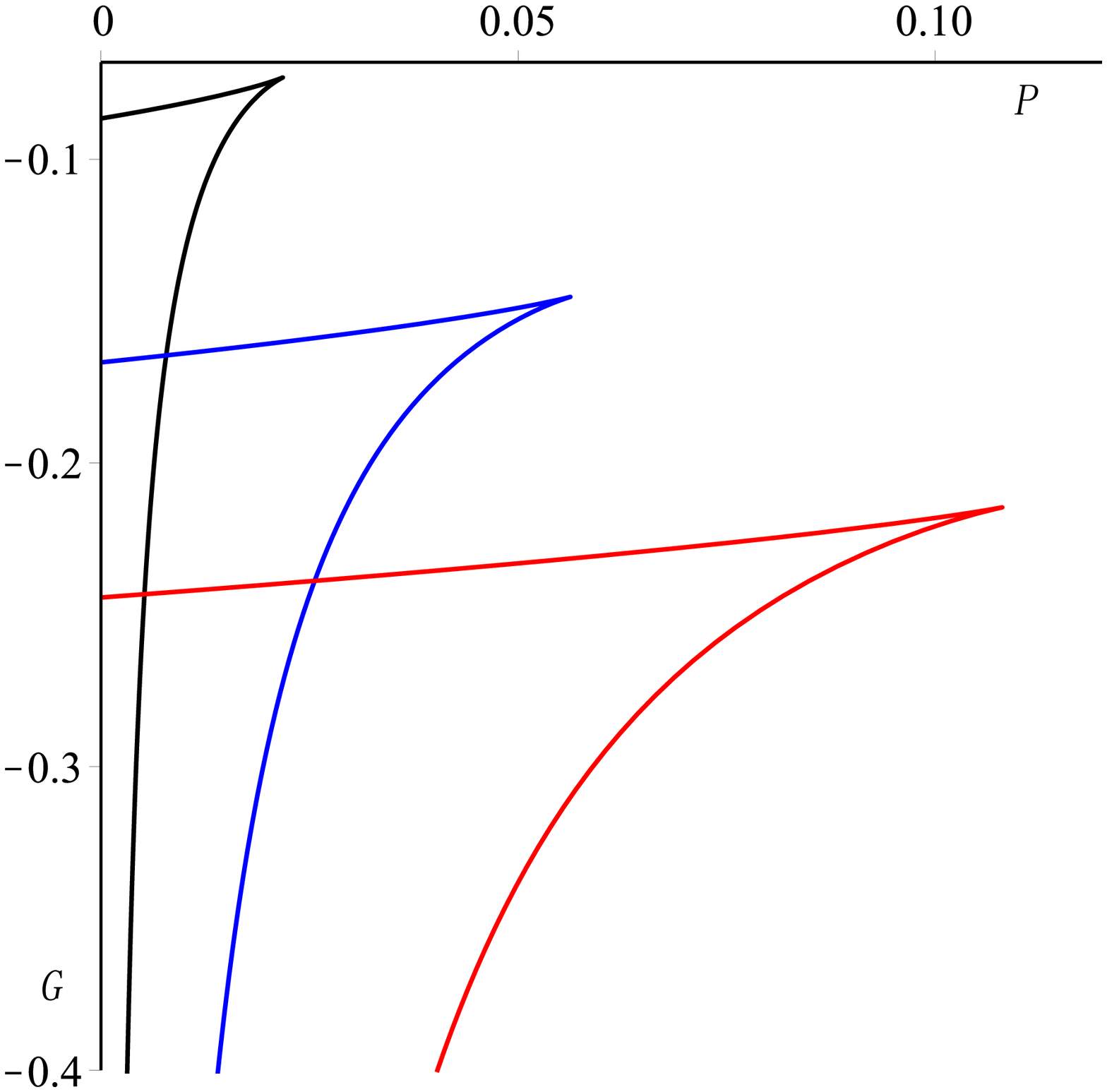}
\caption{The isothermal curves in $P-V$ plane and $G-P$ plane for $(P_2,V_2)$ in $c_1<0$.}
\label{fig7}
\end{center}
\end{figure}

Then we consider the case of $(P_2,V_2)$. Inserting the formulas of $P_2$ and $V_2$ into $f(r_0)$ and $S$, we have

\begin{align}
  &c_0=r_0+\sqrt{1+4\pi T r_0^2},\\
  &S=\frac{1}{3}+\frac{4\pi Tr_0}{3}+\frac{1}{3}\sqrt{1+4\pi T r_0^2},
\end{align}
which are both positive in $r_0>0$. In Fig.\ref{fig7} we present the isotherms in $P-V$ and $G-P$ planes of $(P_2,V_2)$. The black hole also has an analogy with the Hawking-Page phase transition.

In this paper we explore the thermodynamic phase structure for the conformal gravity. Special emphasis is put on the dependence on the parameter $c_1$ from linear-$r$ term in the metric. The case of charged black hole is not considered. In Einstein gravity, the electric charge plays an important role in the thermodynamic phase structure. Although conformal gravity admits charged black hole, the metric does not contain the $\frac{Q^2}{r^2}$ term of the RN black hole \cite{Li:2012gh}, and the \eqref{relation} becomes
\begin{equation}
c_0^2=1+Q^2+3c_1d.
\end{equation}
Thus the qualitative features of the phase structure will remain unchanged.

Due to the rich thermodynamic phase structure of conformal gravity, it would be interesting to investigate its holographic duality. For example, in Einstein gravity we also have the first law about entanglement entropy, which states that the increase of holographic entanglement entropy of a subregion on the AdS boundary is proportional to the increase of the subsystem energy \cite{Bhattacharya:2012mi,Sun:2016til,Lokhande:2017jik,McCarthy:2017amh}. It is worth checking whether this first law is still valid in conformal gravity. Holographic thermalization \cite{Balasubramanian:2011ur,Liu:2013qca} and entanglement growth during the phase transition \cite{Xu:2017wvu} can also be investigated by triggering a sudden local or non-local holographic quench. Wormholes can also be constructed in conformal gravity \cite{Hohmann:2018shl}. In this paper we focus on the AdS space-time; it is necessary to extend the analysis to de Sitter space-time, which shall be addressed in the near future.

\begin{acknowledgments}

Hao Xu would like to thank Yuan Sun and Liu Zhao for useful discussions. We acknowledge the  support by Natural Science Foundation of Guangdong Province (2017B030308003) and the Guangdong Innovative and Entrepreneurial Research Team Program (No.2016ZT06D348), and the Science Technology and Innovation Commission of Shenzhen Municipality (ZDSYS20170303165926217, JCYJ20170412152620376).

\end{acknowledgments}

\providecommand{\href}[2]{#2}\begingroup\raggedright\endgroup

\end{document}